\documentclass[11pt,twoside]{article}
\usepackage{asp2010}

\resetcounters
\def\note #1]{{\bf #1]}}
\def\titnt #1].{}
\def\dd{{\rm d}}
\def\bolddelta{\delta\kern-0.45em\delta\kern-0.45em\delta}
\def\boldr{\mbox{\boldmath$r$}}
\def\bolddelr{\bolddelta \boldr}
\def\muHz{\,\mu{\rm Hz}}
\def\div{{\rm div}\,}
\def\fig{.}
\def\CD{{\cal D}}
\def\CG{{\cal G}}
\def\omegaac{\omega_{\rm ac}}
\def\muHz{\,\mu{\rm Hz}}

\aspvoltitle{Proceedings of the 61st Fujihara Seminar:
Progress in solar/stellar physics \\
with helio- and asteroseismology}
\aspvolauthor{H. Shibahashi, ed.}

\begin{document}

\title{{\it Kepler} asteroseismology of red-giant stars}
\author{J{\o}rgen Christensen-Dalsgaard$^{1,2}$
\affil{$^1$Department of Physics and Astronomy, Aarhus University, 
Ny Munkegade, 8000 Aarhus C}
\affil{$^2$High Altitude Observatory,
National Center for Atmospheric Research,\\
P.O. Box 3000, Boulder, CO 80307, USA}}

\begin{abstract}
The {\it Kepler} mission, launched in March 2009, has revolutionized 
asteroseismology, providing detailed observations of thousands of stars.
This has allowed in-depth analysis of stars ranging from compact hot subdwarfs
to red giants, and including the detection of solar-like oscillations in
hundreds of stars on or near the main sequence.
Here I mainly consider solar-like oscillations in red giants, where
{\it Kepler} observations are yielding results of a perhaps unexpected
richness.
In addition to giving a brief over\-view of the observational and numerical
results for these stars, I present a simple analysis which captures some
of the properties of the observed frequencies.
\end{abstract}

\section{Introduction}

%
\label{sec:intro}
High-precision photometry from space is emerging as an extremely important
astrophysical tool, particularly in two areas: asteroseismology and
the search for extra-solar planets using the transit technique.
The small Canadian MOST satellite \citep[e.g.,][]{Walker2003} has shown
the way, providing very interesting results on stellar oscillations but
also, remarkably, detecting the transit of a planet of just twice
the size of the Earth in orbit around a star of magnitude 6
\citep{Winn2011}.
The CoRoT mission \citep[e.g.,][]{Baglin2009}
was designed with the dual purpose of asteroseismology and
exo-planet research.
The mission, launched in late 2006, has been highly successful in both areas.
The results include the detection of solar-like oscillations in several
main-sequence stars \citep[e.g.,][]{Michel2008} and the detection
of a `super-earth' with a radius around $1.7 \,R_\oplus$ and a mass
around $5 \,M_\oplus$ \citep[e.g.,][]{Leger2009, Queloz2009}.

The {\it Kepler} mission \citep{Boruck2009} uses a telescope 
with a diameter more that three times as big as that of CoRoT.
Also, unlike CoRoT which is in a low-Earth orbit, {\it Kepler}
is in an Earth-trailing heliocentric orbit and hence is able to
observe the same field almost continuously throughout the mission.
The main goal of the mission is
to characterize extra-solar planets, particularly planets of
roughly earth-size in the habitable zone.
This is done through photometric detection of transits of a planet
in front of the host star.
The transit detections of exo-planet candidates
must be followed up by extensive additional analyses and
observations to eliminate possible false positives \citep[e.g.,][]{Brown2003}.
To ensure satisfactory detection statistics {\it Kepler} continuously
observes around 150,000 stars in a field of 110 square degrees in the
Cygnus-Lyra region.
Further details on the spacecraft and mission were provided by
\citet{Koch2010}.

The mission was launched in March 2009 and has been spectacularly
successful.
Amongst the remarkable results have been the statistics of a large number
of exo-planet candidates \citep{Boruck2011},
the detection of the first confirmed rocky exo-planet
\citep{Batalh2011} and the detection of a system with 6 planets,
with masses determined from the planet-planet interactions
\citep{Lissau2011}.

The photometric requirements of the detection of the transit of an Earth-size
planet in front of a star like the Sun, with a reduction in intensity
of around $10^{-4}$, make the mission ideally suited for asteroseismology.
This has led to the establishment of the {\it Kepler} Asteroseismic
Investigation \citep[KAI,][]{Christ2008}.
Most objects are observed at a cadence of around 30 min, but a subset of
up to 512 stars, the selection of which can be changed on a monthly basis,
are observed at a one-minute cadence.
A large fraction of these short-cadence slots have been made available
to asteroseismology of stars near the main sequence, while the
long-cadence data can be used for asteroseismology of larger and more
evolved stars, particularly red giants.
To make full use of the huge amount of data from the mission,
the Kepler Asteroseismic Science Consortium (KASC)%
\footnote{see {\tt http://astro.phys.au.dk/KASC/}.}
has been set up \citep{Kjelds2010};
at the time of writing (September 2011) this has more than 500
members, organized into 13 working groups dealing with different types
of objects.
The asteroseismic data are made available to the members of the KASC through
the Kepler Asteroseismic Science Operations Centre (KASOC) in Aarhus,
which also deals with the management and internal refereeing of the
papers resulting from the KAI.
In the early part of the mission a large number of stars were observed
for typically one month each, in a survey phase designed to characterize
the properties of stars in the field and select targets for the
now ongoing detailed studies of specific objects.

The analysis of the asteroseismic data benefits greatly from
additional information about the stars.
Basic data for a very large number of stars in the {\it Kepler}
field, obtained from multi-colour photometry, are available in
the Kepler Input Catalogue \citep[KIC,][]{Brown2011}.
However, extensive efforts are under way to secure more accurate data
for selected objects \citep[e.g.,][]{Molend2010, Molend2011, Uytter2010}.


A review of the early {\it Kepler} asteroseismic results was given
by \citet{Gillil2010}, while more recent reviews have been
provided by \citet{ChristT2011} and \citet{Christ2011a, Christ2011b}.
Here I give a very brief over\-view of some of the key findings;
also, I discuss in somewhat more detail {\it Kepler} asteroseismology 
of red giants, which has turned out to be perhaps the most fascinating area.

\section{Properties of stellar oscillations}

%
\label{sec:oscprop}
As a background for the discussion of individual types of stars below,
it is probably useful to provide a brief overview of the properties
of stellar oscillations.
More detailed expositions have been provided, for example,
by \citet{Unno1989}, \citet{Christ2004}, \citet{Aerts2010} and
\citet{Christ2011a}.

We consider only small-amplitude adiabatic oscillations
of spherically symmetric stars.
The dependence of a mode of oscillation on co-latitude $\theta$ and
longitude $\phi$ can be written as a spherical harmonic $Y_l^m(\theta, \phi)$,
where the degree $l$ provides a measure of the total number of nodal
lines on the stellar surface and the azimuthal order $m$ measures the
number of nodal lines crossing the equator.
In the absence of rotation and other departures from spherical symmetry
the frequencies are independent of $m$.
From a physical point of view there are two dominant restoring forces
at work: pressure perturbations and gravity working on density perturbations.
The former case dominates in acoustic waves and leads to modes of
oscillation characterized as p modes, while the latter corresponds to
internal gravity waves and leads to modes characterized as g modes.

A very good understanding of the properties of the modes can be obtained
from simple asymptotic analysis which in fact turns out to have 
surprisingly broad applicability in the study of asteroseismically
interesting stars.
By generalizing an analysis by \citet{Lamb1909},
\citet{Deubne1984} showed that the oscillations approximately satisfy 
\begin{equation}
{\dd^2 X \over \dd r^2} = -K(r) X \; ;
\label{eq:asymp}
\end{equation}
here $r$ is distance to the centre and
$X = c^2 \rho^{1/2} \div \bolddelr$, where $\bolddelr$
is the displacement vector, $c$ is the adiabatic sound speed and 
$\rho$ is density.
Also
\begin{equation}
K = {1 \over c^2} \left[ S_l^2 \left({N^2 \over \omega^2} - 1\right)
+ \omega^2 - \omegaac^2 \right] \; ,
\label{eq:kasymp}
\end{equation}
$\omega$ being the frequency of oscillation,
is determined by three characteristic frequencies of the star:
the {\it Lamb frequency} $S_l$, with
\begin{equation}
S_l^2 = {l(l+1) c^2 \over r^2} \; ,
\label{eq:lamb}
\end{equation}
the {\it buoyancy frequency} (or {\it Brunt-V\"ais\"al\"a frequency}) $N$,%
\footnote{Note the relation of $N$ to convective instability: in
convectively unstable regions $N^2 < 0$.}
\begin{equation} 
N^2 = g \left( {1 \over \Gamma_1} {\dd \ln p \over \dd r} 
- {\dd \ln \rho \over \dd r} \right) \; ,
\label{eq:buoy}
\end{equation}
where $g$ is the local gravitational acceleration,
and the {\it acoustic cut-off frequency} $\omegaac$,
\begin{equation}
\omegaac^2 = {c^2 \over 4 H^2} \left(1 - 2 {\dd H \over \dd r} \right)
 \; ,
\end{equation}
where $H = - (\dd \ln \rho / \dd r)^{-1}$ is the density scale height.
The properties of a mode are largely determined by the regions in the
star where the eigenfunction oscillates as a function of $r$, i.e.,
where $K > 0$.
In regions where $K < 0$ the eigenfunction locally increases or
decreases exponentially with $r$.
There may be several oscillatory regions, but typically the amplitude is
substantially larger in one of these than in the rest, defining the
region where the mode is said to be trapped.

The acoustic cut-off frequency is generally large near the stellar surface
and small in the interior.
For the low-degree modes relevant here
the term in $S_l^2$ in Eq.~(\ref{eq:kasymp}) is small near the surface,
and the properties of the oscillations are determined by the
magnitude of $\omega$ relative to the atmospheric value of $\omegaac$:
when $\omega$ is less than $\omegaac$ the eigenfunction decreases 
exponentially in the atmosphere, and the mode is trapped in 
the stellar interior;
otherwise, the mode is strongly damped by the loss of energy through running 
waves in the atmosphere.

The properties of the oscillations in the stellar interior are controlled
by the behaviour of $S_l$ and $N$ (see Fig.~\ref{fig:charfreq}).
In unevolved stars $N$ is typically small compared with the characteristic
frequencies of p modes.
For these, therefore, $K \simeq (\omega^2 - S_l^2)/c^2$
(neglecting $\omegaac$), and the modes are trapped in the region
where $\omega > S_l$ in the outer parts of the star, with a lower turning
point, $r = r_{\rm t}$, such that
\begin{equation}
{c(r_{\rm t}) \over r_{\rm t}} = {\omega \over \sqrt{l(l+1)}} \; .
\end{equation}
At low degree the cyclic frequencies of p modes approximately satisfy 
\begin{equation}
\nu_{nl} = {\omega_{nl} \over 2 \pi} \simeq
\Delta \nu \left(n + {l \over 2} + \epsilon \right) - d_{nl} \; 
\label{eq:pasymp}
\end{equation}
\citep{Vandak1967, Tassou1980, Gough1993};
here $n$ is the radial order of the mode,
\begin{equation}
\Delta \nu = \left(2 \int_0^R {\dd r \over c} \right)^{-1}
\label{eq:delnu}
\end{equation}
is the inverse sound travel time
across a stellar diameter, $R$ being the surface radius,
$\epsilon$ is a frequency-dependent phase
that reflects the behaviour of $\omegaac$ near the stellar surface
and $d_{nl}$ is a small correction that in main-sequence stars
predominantly depends on the sound-speed gradient in the stellar core.
On the other hand, g modes have frequencies below $N$ and typically
such that $\omega \ll S_l$ in the relevant part of the star;
in this case the modes are trapped in a region defined by $\omega < N$.
Here the oscillation {\it periods} satisfy a simple asymptotic relation:
\begin{equation}
\Pi_{nl} = {2 \pi \over \omega_{nl}} 
\simeq \Delta \Pi_l (n + \epsilon_{\rm g}) \; 
\label{eq:gasymp}
\end{equation}
\citep{Vandak1967, Tassou1980},
where $\epsilon_{\rm g}$ is a phase and
\begin{equation}
\Delta \Pi_l = {2 \pi^2 \over \sqrt{l(l+1)}}
\left(\int_{r_1}^{r_2} N {\dd r \over r} \right)^{-1} \; ,
\label{eq:gperspac}
\end{equation}
the integral being over the region where the modes are trapped.

As discussed in Section~\ref{sec:redgiant} (see also Bedding, these proceedings)
the situation is considerably more complicated in evolved stars with a compact
core;
this leads to a high value of $g$ and hence $N$ in the interior of the
star, such that modes may have a g-mode character in the deep interior,
and a p-mode character in the outer, parts of the star.
Such {\it mixed modes} have a very substantial diagnostic potential.

\section{Solar-like oscillations}

\label{sec:solarlike}
The solar oscillations are most likely intrinsically damped
\citep{Balmfo1992} and excited stochastically by the turbulent
near-surface convection \citep[e.g.,][]{Goldre1977}.
Thus such oscillations can be expected in all stars with significant
outer convection zones \citep{Christ1983, Houdek1999}.
An interesting example is the very recent detection, based on {\it Kepler} data,
of solar-like oscillations in a $\delta$ Scuti star \citep{Antoci2011}.
The combination of damping and excitation leads to a well-defined,
bell-shaped envelope of power with a maximum at a cyclic frequency 
$\nu_{\rm max} = \omega_{\rm max}/2 \pi$ which approximately scales as the
acoustic cut-off frequency $\omegaac$ in the atmosphere \citep{Brown1991}.
Assuming adiabatic oscillations in an isothermal atmosphere, this yields
\begin{equation}
\nu_{\rm max} \propto \omegaac \propto M R^{-2} T_{\rm eff}^{-1/2} \; ,
\label{eq:numax}
\end{equation}
where $M$ is the mass, and $T_{\rm eff}$ the effective temperature, of the star.
This scaling has substantial observational support
\citep[e.g.,][]{Beddin2003, Stello2008}, while it is still not fully
understood from a theoretical point of view \citep[but see][]{Belkac2011}.
For stars on or near the main sequence the frequencies approximately
satisfy the asymptotic relation (\ref{eq:pasymp}).
Here the large frequency separation roughly scales, in accordance
with homology scaling, as the square root of the mean stellar density, i.e.,
\begin{equation}
\Delta \nu \propto M^{1/2} R^{-3/2} \; ,
\label{eq:dnuscale}
\end{equation}
although with some departures from strict homology that largely depend
on $T_{\rm eff}$ \citep{White2011}.
A detailed discussion of the observational properties of
solar-like oscillations was provided by \citet{Beddin2011}.


In the early phases of the {\it Kepler} project a survey 
of potential asteroseismic targets was carried out, each typically observed for
one month.
This led to the detection of a huge number of main-sequence and subgiant 
stars showing solar-like oscillations, increasing the number of the known
cases by more than a factor of 20, relative to earlier 
ground-based observations and the observations with the CoRoT mission.
As discussed by \citet{Chapli2011a} this provides information about basic 
stellar properties,
through the application of the scaling laws
for $\nu_{\rm max}$ and $\Delta \nu$ 
(Eqs~\ref{eq:numax} and \ref{eq:dnuscale}),
yielding a very interesting overview of the distribution, 
in mass and radius, of stars in the solar neighbourhood.
These extensive data also allow a calibration of scaling relations
that can be used to predict the detectability of solar-like oscillations
for a given star \citep{Chapli2011b};
this is important for the scheduling of further {\it Kepler} observations and
will, suitably generalized, be very valuable for the definition of
other observing campaigns. 

A selected set of these stars are now being observed for much longer, to 
improve the precision and level of detail of the observed frequencies and
hence allow detailed characterization of the stellar properties and 
internal structure, through fits to the individual observed frequencies.
Interesting examples of such analyses were carried out by
\citet{Metcal2010} and \citet{DiMaur2011} (see also Di Mauro et al.,
these proceedings), in both cases for stars
evolved somewhat beyond the main sequence where modes of mixed p- and g-mode
character provided more sensitive information about the stellar properties
(see also Bedding, these proceedings).
A very interesting analysis of the diagnostic potentials of mixed modes,
based on CoRoT observations, was presented by \citet{Deheuv2011}.


An important aspect of the asteroseismology of solar-like stars is the
application to stars found by {\it Kepler} to be planet hosts.
Here the analysis allows the precise determination of the stellar mass
and radius, essential for characterizing the properties of the planets
detected through the transit technique and follow-up radial velocity
measurements,
and determination of the stellar age through model fits to the observed
frequencies provides a measure of the age of the planetary system.
An early example of such analyses, for the previously known planet host
HAT-P-7, was provided by \citet{Christ2010}, and asteroseismic results
were important for the characterization of the first rocky planet
detected outside the solar system, Kepler-10b \citep{Batalh2011}.



\section{Asteroseismology of red giants}

\label{sec:redgiant}
Perhaps the most striking aspect of {\it Kepler} asteroseismology has
been the study of solar-like oscillations in red-giants stars.
These have evolved beyond the phase of central hydrogen burning,
developing a helium core surrounded by a thin region, the hydrogen-burning 
shell, providing the energy output of the star.
The core contracts while the outer layers expand greatly,
accompanied by a reduction of the effective temperature,
until the star reaches and evolves up the Hayashi track, at nearly constant
$T_{\rm eff}$. 
At the tip of the red-giant branch the temperature in the core reaches around
100\,MK, enough to start efficient helium burning.
Helium ignition is followed by a readjustment of the internal structure, 
leading to some reduction in the surface radius, 
and maintaining the hydrogen-burning shell as a substantial contributor to
the total energy output of the star.
This central helium-burning phase is rather long-lived, with the stars
changing relatively little;
in stellar clusters this leads to an accumulation of stars in this phase,
which is therefore known as the `clump phase'.

\begin{figure}[htb]
\begin{center}
\includegraphics[width=9cm]{\fig/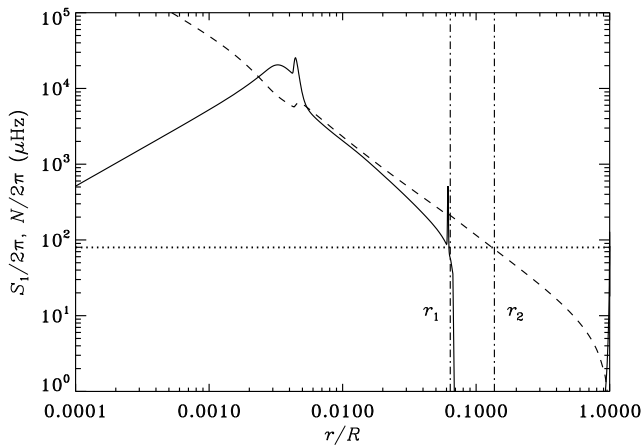}
\end{center}
\vskip -0.8cm
\caption{Lamb frequency $S_l$ for $l = 1$ (dashed line)
and buoyancy frequency $N$ (solid line), against fractional radius,
in a $1\,{\rm M}_\odot$ red-giant model with heavy-element abundance
$Z = 0.02$, radius $7\,{\rm R}_\odot$ and luminosity $18.4\,{\rm L}_\odot$.
The horizontal dotted line is at $0.7 \omegaac(R)/2 \pi$,
corresponding roughly to the predicted frequency $\nu_{\rm max}$ of maximum 
power.
The vertical dot-dashed lines mark the outer limit $r_1$ of the g-mode,
and the lower limit $r_2$ of the p-mode, propagating regions at this frequency.
\label{fig:charfreq}
}
\end{figure}

A red giant has a very compact core, containing a substantial fraction of the
stellar mass, and a very extensive convective envelope.
This is reflected in the characteristic frequencies,
shown in Fig.~\ref{fig:charfreq} for a typical red-giant model.
As is clear from the behaviour of the buoyancy frequency $N$, the outer 
convective envelope extends over more than 90 \% of the stellar radius.
The border of the helium core, with a mass of $0.21 \, M$,
is marked by the small local maximum in $N$ near $r = 0.0044 R$; 
the large mass within a small region
leads to a very large local gravitational acceleration,
causing the huge buoyancy frequency (cf.\ Eq.~\ref{eq:buoy}).
At a typical oscillation frequency, marked by the horizontal dotted line,
the mode behaves as a p mode in the region outside $r = r_2$, marking
the bottom of the outer acoustic cavity, and as a g mode inside $r = r_1$.
This causes a mixed character of the modes
(see also Bedding, these proceedings).

\subsection{Properties of red-giant oscillations}

Early predictions of solar-like oscillations in red giants were made
by \citet{Christ1983}.
This was followed by detections in ground-based observations
\citep[e.g.,][]{Frands2002};
however, it was only with the CoRoT observations by \citet{DeRidd2009} that
the presence of nonradial solar-like oscillations in red giants was definitely
established.
Very extensive observations of large numbers of stars
have been made with both CoRoT and {\it Kepler}
\citep[e.g.,][]{Hekker2009, Beddin2010, Mosser2010, Kallin2010, Hekker2011}.
This has allowed the determination of global stellar properties from fits
to the large frequency separation $\Delta \nu$ (cf.\ Eq. \ref{eq:pasymp})
and the frequency
$\nu_{\rm max}$ at maximum power, possibly supplemented by additional
observations in analyses based on stellar model grids,
allowing population studies of the red giants 
\citep[e.g.,][]{Miglio2009, Miglio2011}.
Particularly interesting are the possibilities for studying the open clusters
in the {\it Kepler} field \citep{Stello2010, Stello2011a, Basu2011}.
The large number and variety of stars observed have allowed the study of
the overall properties of the oscillation frequencies, along the lines
of the p-mode asymptotic relation (Eq.~\ref{eq:pasymp}),
\citep{Huber2010, Mosser2011a}, in what the latter call the universal pattern
of the frequencies, and the diagnostic potential of this pattern
\citep{Montal2010}.
Similarly, it has been possible to test scaling relations for mode properties,
including their amplitudes, over a broad range of stellar parameters
\citep{Huber2011, Mosser2011b, Stello2011b}.
\citet{DiMaur2011} demonstrated the diagnostic potential of
analysing the individual frequencies in a relatively unevolved red giant,
while \citet{Jiang2011} carried out a detailed analysis of a somewhat
more evolved star.


\begin{figure}[htb]
\begin{center}
\includegraphics[width=9cm]{\fig/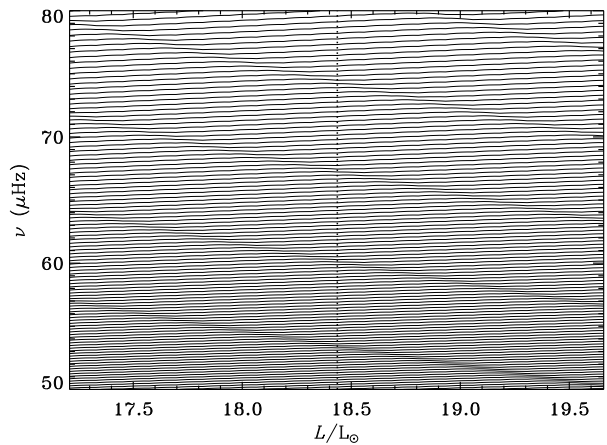}
\end{center}
\vskip -0.8cm
\caption{
Computed frequencies for a short segment of a $1\,{\rm M}_\odot$ evolution
sequence, plotted against luminosity in solar units.
The age goes from 11.765 to 11.777\,Gyr.
The vertical dotted line indicates the model illustrated 
in Figs~\ref{fig:charfreq}, \ref{fig:inertia} and~\ref{fig:perspac}.
\label{fig:freqs}
}
\end{figure}

To understand the observed oscillation spectra it is informative to consider
in more detail the oscillation properties of red-giant models;
here I consider models in the vicinity of the model illustrated in
Fig.~\ref{fig:charfreq}.
The huge buoyancy frequency in the core of the model leads to very small
g-mode period spacings (cf.\ Eq.~\ref{eq:gperspac}):
for $l = 1$ the result is $\Delta \Pi_1 = 1.23$\,min.
The full range of radial modes, up to the acoustical cut-off frequency,
extends in cyclic frequency from 11 to $107 \muHz$.
In that interval we therefore expect around 1100 g modes of degree $l = 1$,
the number scaling like $\sqrt{l(l+1)/2}$ for higher degree.%
\footnote{This also places heavy constraints on the numerical precision
of the calculation of the oscillations.
In the numerical calculations illustrated here
I used 19,200 mesh points, distributed to match the 
asymptotic properties of the eigenfunctions, which secured reasonable
precision of the results (Christensen-Dalsgaard et al., in preparation).}

Most of these modes have their highest amplitude in the core of the model
and hence predominantly have the character of g modes.
However, there are frequencies where the eigenfunctions decrease with 
increasing depth in the region between $r_2$ and $r_1$ 
(cf.\ Fig.~\ref{fig:charfreq}).
These frequencies define {\it acoustic resonances} where the modes have 
their largest amplitude in the outer acoustic cavity and the modes 
predominantly have the character of p modes.

The properties of the frequencies are illustrated in Fig.~\ref{fig:freqs},
showing the evolution of $l = 1$ modes with age,
in a relatively limited range in frequency.
The extremely high density of modes is evident,
as are the branches of acoustic resonances, with frequency decreasing
roughly as $R^{-3/2}$, proportional to the mean density of the star.
As discussed by Bedding (these proceedings) these undergo avoided crossings 
with the g modes, whose frequencies increase somewhat with age as the buoyancy
frequency in the core increases.
The acoustic resonances, together with the frequencies of radial modes,
approximately satisfy the asymptotic relation given in Eq.~(\ref{eq:pasymp}).

\begin{figure}[htb]
\begin{center}
\includegraphics[width=9cm]{\fig/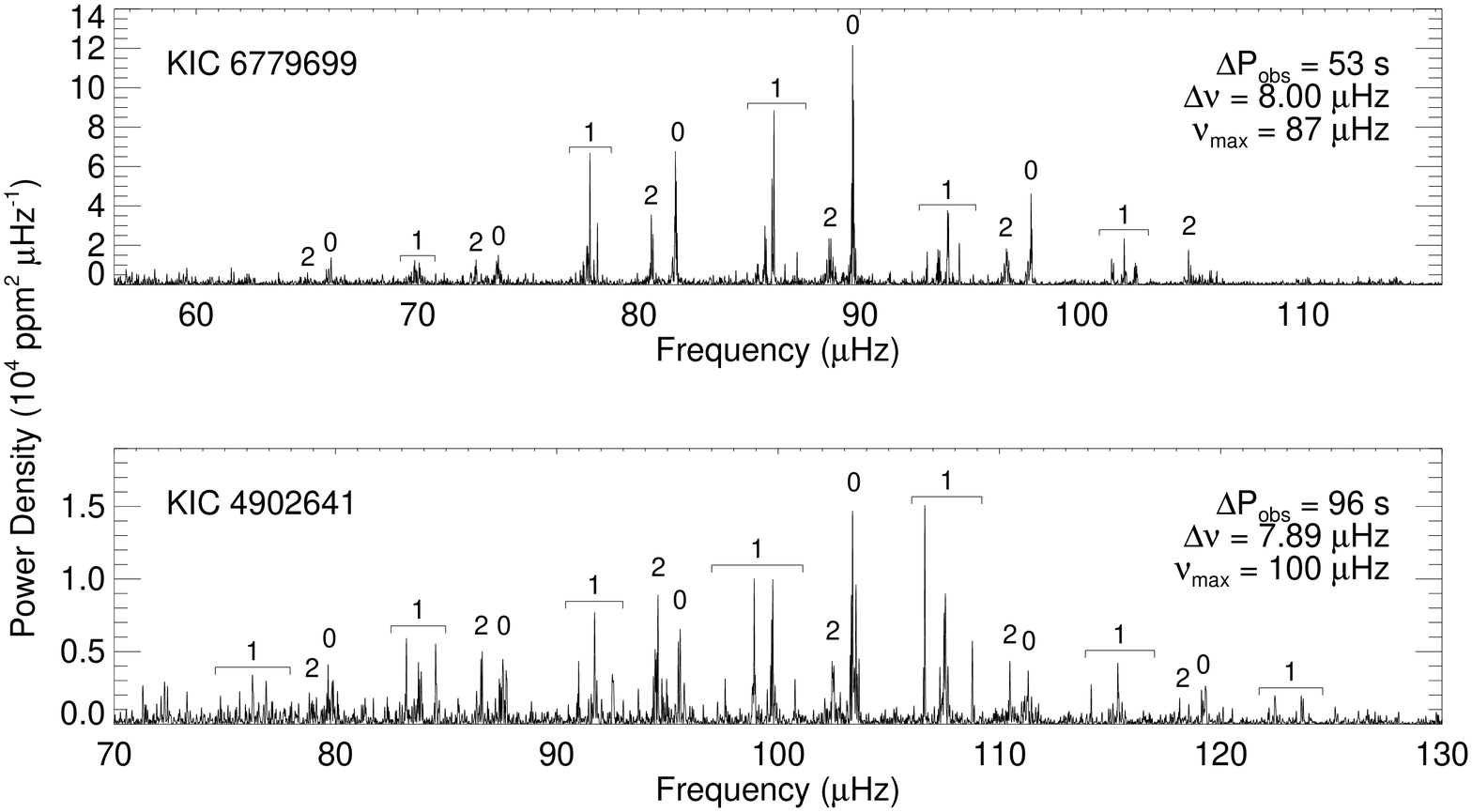}
\end{center}
\vskip -0.8cm
\caption{
Power spectra of two red giants observed by {\it Kepler} for 13 and 10
months, respectively; the stars are identified by
the Kepler Input Catalog (KIC) numbers. 
As indicated in the figure, the stars have similar $\Delta \nu$ 
and $\nu_{\rm max}$ and hence similar overall properties. 
The horizontal bars marked `1' indicate the dipole forests from which
were inferred
rather different observed period spacings, $\Delta P_{\rm obs}$,
of 52 and 96\,s for the upper and lower spectrum.
Further analysis determined the asymptotic g-mode spacings as 74 and 147\,s,
respectively.
KIC\,6779699 is in the ascending red-giant phase, whereas
KIC\,4902641 is in the clump phase, with central helium burning.
Adapted from \citet{Beddinetal2011}.
\label{fig:obsspec}
}
\end{figure}

A major breakthrough in the asteroseismic study of red giants was the detection
of 'dipole forests' at the acoustic resonances, i.e., groups of $l = 1$
modes excited to observable amplitudes
\citep{Beck2011, Beddinetal2011, Mosser2011c}.
These showed an approximately uniform period spacing, as expected for 
g modes from Eq.~(\ref{eq:gasymp}),
and in some cases it was possible from the observed
spacings to extrapolate to the period spacing $\Delta \Pi_1$ for the pure
g modes.
Strikingly, \citet{Beddinetal2011} demonstrated a substantial difference in
the period spacing between stars in the shell hydrogen burning phase
and `clump stars' which in addition had core helium burning.
This is illustrated in the observed spectra shown 
in Fig.~\ref{fig:obsspec}.
It was argued by \citet{Christ2011a} that the larger period spacing in
the helium-burning stars is a natural consequence 
of the very temperature-dependent energy generation near the centre 
in these stars: this leads to a convective core which restricts
the range of the integral in Eq.~(\ref{eq:gperspac}), thus reducing its
value and hence increasing $\Delta \Pi$.

\begin{figure}[htb]
\begin{center}
\includegraphics[width=9cm]{\fig/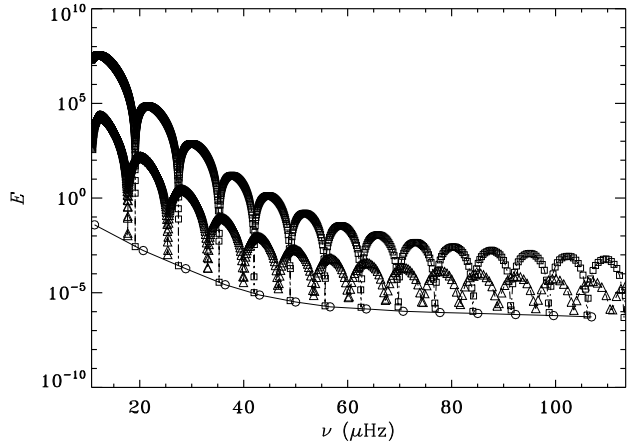}
\end{center}
\vskip -0.8cm
\caption{
Mode inertias (cf.\ Eq.~\ref{eq:inertia})
for the $1\,{\rm M}_\odot$ model illustrated in Fig.~\ref{fig:charfreq}.
Radial modes, with $l = 0$, are shown with circles connected by a line,
$l = 1$ modes are shown with triangles and $l = 2$ modes with squares.
\label{fig:inertia}
}
\end{figure}

\begin{figure}[htb]
\begin{center}
\includegraphics[width=9cm]{\fig/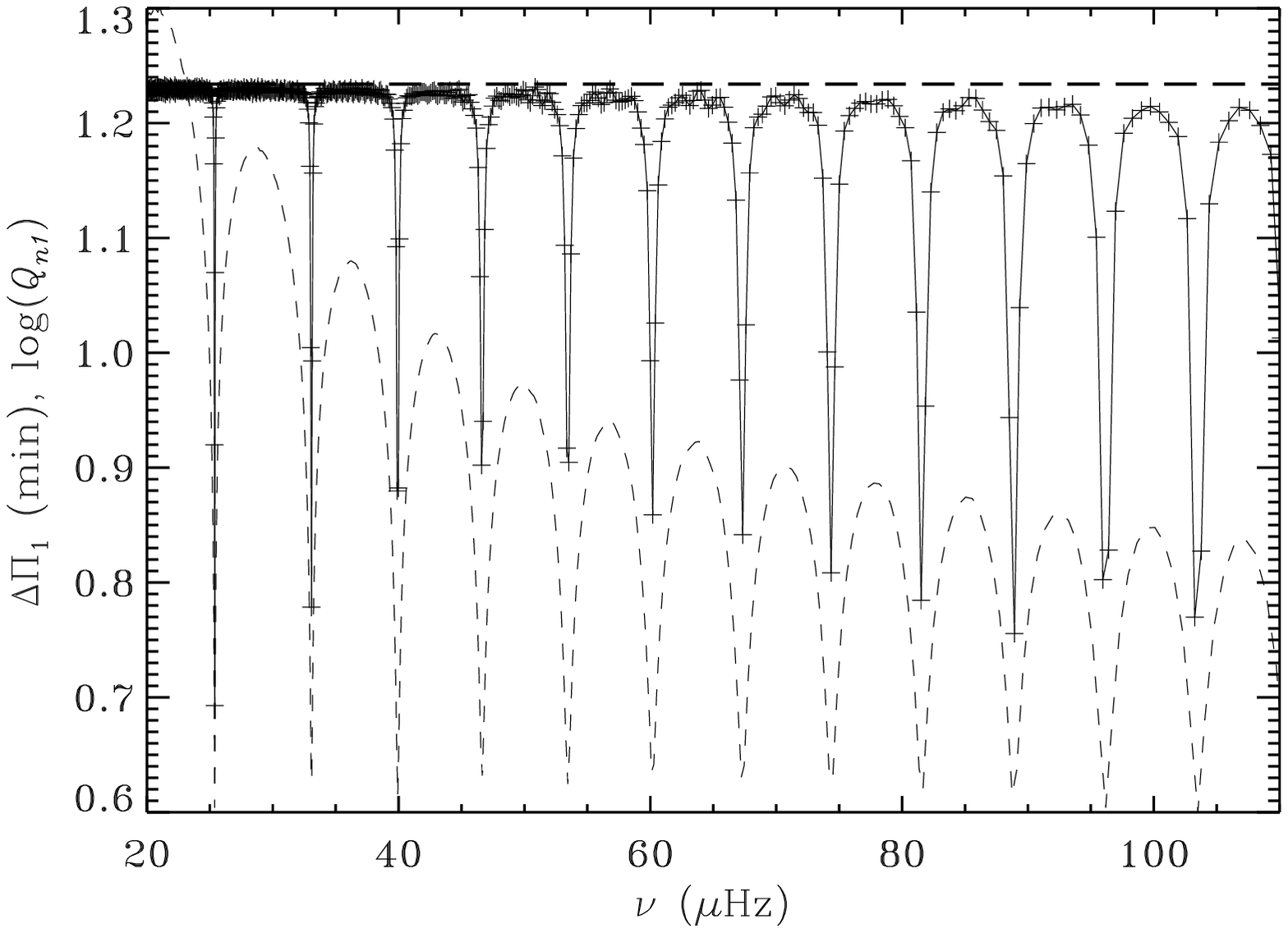}
\end{center}
\vskip -0.8cm
\caption{
The pluses connected by solid lines show period spacings for modes with $l = 1$
in the $1\,{\rm M}_\odot$ model illustrated in Fig.~\ref{fig:charfreq}.
The heavy horizontal dashed line shows the asymptotic period spacing
(cf.\ Eq.~\ref{eq:gperspac}).
The thin dashed curve shows, on an arbitrary scale, the logarithm of
the mode inertia divided by the inertia of radial modes (see text).
\label{fig:perspac}
}
\end{figure}

An important quantity characterizing the properties of the modes is 
the normalized mode inertia
\begin{equation}
E = {\int_V \rho |\bolddelr|^2 \dd V \over M |\bolddelr_{\rm s}|^2 } \; ,
\label{eq:inertia}
\end{equation}
where $\bolddelr_{\rm s}$ is the
surface displacement, and the integral is over the volume $V$ of the star.
Modes of predominantly acoustic nature, including the radial modes, have
their largest displacement in the outer parts of the star and hence a relatively
low inertia, whereas g-dominated modes have a large amplitude in the deep
interior and hence a high inertia.
This is illustrated in Fig.~\ref{fig:inertia} for modes of degree
$l = 0$, 1 and 2.
The acoustic resonances, the inertia decreasing to close to the value for
a radial mode of corresponding frequency, are evident, while for the 
intermediate g-dominated modes the inertia is obviously much higher.


On the assumption of damped modes excited stochastically by near-surface
convection, the mode inertia is important for understanding the observed
amplitudes.
A very illuminating discussion of this was provided by \citet{Dupret2009}.
In the frequent case where the damping is predominantly near the surface,
the mode lifetime is proportional to $E$, with the g-dominated modes
having lifetimes of years.
Also, the mean square amplitude is inversely proportional to $E$.
However, the visibility of the modes in the power spectrum is determined
with the {\it peak height} which under these circumstances is
independent of $E$, assuming observations extending for substantially more than
the mode lifetimes \citep[see also][]{Chapli2005}.
The present observations are much shorter than the typical lifetimes of the 
g-dominated modes but still allow several modes to be seen close to 
the acoustic resonances,
particularly for $l = 1$, leading to the dipolar forests of peaks.

As noted by \citet{Beck2011} and \citet{Beddinetal2011}
the signature of modes affected by the g-mode behaviour is an approximately
uniform period spacing.
This is illustrated for the model in Fig.~\ref{fig:perspac}.
For the bulk of the modes, which are predominantly g modes, the spacing
is close to the asymptotic value, indicated by the horizontal dashed line.
Near the acoustic resonances the avoided crossings cause a reduction
in the period spacing, of a characteristic 'V'-shaped pattern.
This is closely related to the behaviour of the mode inertia,
illustrated in figure by the scaled mode inertia 
$Q_{nl} = E_{nl}/ \bar E_0(\omega_{nl})$ 
where $\bar E_0(\omega_{nl})$ is the inertia of the radial modes, interpolated
to the frequency $\omega_{nl}$ of the mode considered.

\subsection{A simple asymptotic analysis}

%
To understand the properties of the modes it is useful to illustrate them with
a crude approximation to the asymptotic description provided 
by Eqs~(\ref{eq:asymp}) and (\ref{eq:kasymp}).
We neglect $\omegaac^2$ whose main effect is the ensure reflection of the
mode at the stellar surface; this causes a phase shift which can easily be
included.
From the behaviour of the characteristic frequencies illustrated 
in Fig.~\ref{fig:charfreq}
we can approximately divide the star into three regions, separated at $r = r_1$
and $r_2$ (see figure):

\begin{enumerate}

\item $r < r_1$: here $S_l^2 \gg \omega^2$ and $N^2 \gg \omega^2$, except 
near the turning points.
Here we approximate $K$ by
\begin{equation}
K \simeq {l(l+1) \over r^2} \left( {N^2 \over \omega^2} - 1 \right)
\simeq {l(l+1) \over r^2} {N^2 \over \omega^2} \; ,
\end{equation}
where the last approximation is valid except near the turning points where
$\omega = N$.

\item $r_1 < r < r_2$: here $S_l^2 \gg \omega^2 \gg N^2$.
Thus $K$ is simply approximated by 
\begin{equation}
K \simeq - {l(l+1) \over r^2} \; .
\end{equation}

\item $r_2 < r$: Here $N^2 \ll \omega^2$ and $S_l^2 \ll \omega^2$ except
near the turning point where $\omega = S_l$.
Thus we approximate $K$ by
\begin{equation}
K \simeq {1 \over c^2} \left( \omega^2 - S_l^2 \right) 
\simeq {\omega^2 \over c^2} \; ,
\end{equation}
where the last approximation is valid except near the turning point.

\end{enumerate}

With these approximations we can write down the solutions in the different 
regions based on the JWKB approximation, neglecting, however, the details 
of the behaviour near the turning points.%
\footnote{These can be treated with a proper JWKB analysis
\citep[e.g.,][]{Gough2007}; the effect, however, would essentially just be
to change the phases of the solution, with no qualitative change in 
the behaviour found here.}
In region 1) the solution can be written as
\begin{equation}
X(r) \simeq A^{\rm (g)} (r) \cos (\Psi^{\rm (g)}(r) + \phi^{\rm (g)}) \; ,
\label{eq:gapprox}
\end{equation}
where 
\begin{equation}
\Psi^{\rm (g)} = \int_0^r {L \over r'} {N \over \omega}  \dd r' \; ,
\end{equation}
with $L = \sqrt{l(l+1)}$, and $\phi^{\rm (g)}$ is a phase that depends on 
the behaviour near the centre and the turning points.
In region 2) the approximate solution is simply
\begin{equation}
X(r) \simeq a_+ \left( {r \over r_1} \right)^L
+ a_- \left( {r \over r_1} \right)^{-L} \; ,
\end{equation}
where $a_+$ and $a_-$ are integration constants determined by 
fitting the solutions.
Finally, in region 3) the solution can be written as
\begin{equation}
X(r) \simeq A^{\rm (p)} (r) \cos (\Psi^{\rm (p)}(r) + \phi^{\rm (p)}) \; ,
\label{eq:papprox}
\end{equation}
where 
\begin{equation}
\Psi^{\rm (p)} = \int_r^R \omega {\dd r' \over c} \; ,
\end{equation}
and $\phi^{\rm (p)}$ is a phase that depends on 
the behaviour near the surface (thus including the contribution from
the reflection caused by $\omegaac$) and the turning point.
In Eqs~(\ref{eq:gapprox}) and (\ref{eq:papprox}) the amplitude functions
$A^{\rm (g)}(r)$ and $A^{\rm (p)}(r)$
can be obtained from the JWKB analysis; since they are irrelevant for the
subsequent analysis, I do not present them explicitly.

The full solution is obtained by requiring continuity of $X$ and $\dd X/\dd r$
at $r_1$ and $r_2$.
Neglecting the derivatives of $A^{\rm (g)}(r)$ and $A^{\rm (p)}(r)$
this yields
\begin{eqnarray}
&& \sin(\Psi^{\rm (g)}(r_1) + \phi^{\rm (g)} - \pi/4)
\sin(\Psi^{\rm (p)}(r_2) + \phi^{\rm (p)} - \pi/4) + \nonumber \\
&& \zeta \cos(\Psi^{\rm (g)}(r_1) + \phi^{\rm (g)} - \pi/4)
\cos(\Psi^{\rm (p)}(r_2) + \phi^{\rm (p)} - \pi/4) = 0 \; ,
\label{eq:fulldisp}
\end{eqnarray}
where $\zeta = (r_1/r_2)^{2 L}$ is a measure of the coupling between 
regions 1) and 3).
If $\zeta =0$, so that the regions are completely decoupled, the
eigenfrequencies obviously satisfy
\begin{equation}
\Psi^{\rm (g)}(r_1) = \int_0^{r_1} {L \over r} {N \over \omega}  \dd r 
= n \pi - \tilde \phi^{\rm (g)} \; ,
\label{eq:pureg}
\end{equation}
or 
\begin{equation}
\Psi^{\rm (p)}(r_2) = \int_{r_2}^R \omega {\dd r \over c} 
= k \pi - \tilde \phi^{\rm (p)} \; ,
\label{eq:purep}
\end{equation}
where $n$ and $k$ are integers and 
$\tilde \phi^{\rm (g)} = \phi^{\rm (g)} - \pi/4$,
$\tilde \phi^{\rm (p)} = \phi^{\rm (p)} - \pi/4$.
Equation (\ref{eq:pureg}) leads to Eqs~(\ref{eq:gasymp})
and (\ref{eq:gperspac}),
while Eq.~(\ref{eq:purep}) essentially corresponds to Eq.~(\ref{eq:pasymp}),
taking into account that the behaviour near the lower turning point
leads to the dependence on $l$.%
\footnote{For purely acoustic modes extending over the whole star, this follows
from an analysis of the oscillations near the centre
\citep[e.g.,][]{Gough1993}.
In the present case of a red giant,
where the lower turning point is in a region outside the compact core,
this behaviour is in fact not fully understood.}

\begin{figure}[htb]
\begin{center}
\includegraphics[width=9cm]{\fig/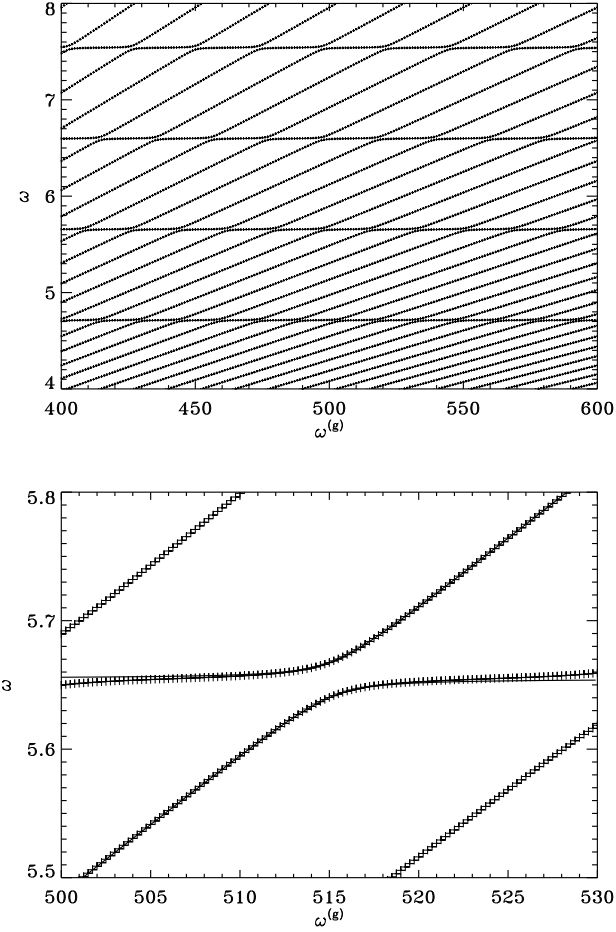}
\end{center}
\vskip -0.8cm
\caption{
The upper panel shows solutions to the dispersion relation in
Eq.~(\ref{eq:fulldisps}), as a function of $\omega^{\rm (g)}$,
keeping $\omega^{\rm (p)} = 0.3$ and $\zeta = 0.01$ fixed.
The lower panel shows a small segment of the solution; 
the solid curves show the approximate solution near the avoided
crossing, given by Eq.~(\ref{eq:avcross}).
\label{fig:avcross}
}
\end{figure}

To analyse the full dispersion relation, Eq.~(\ref{eq:fulldisp}), 
we neglect the dependence of the turning points $r_1$ and $r_2$ on
frequency and write the relation as
\begin{equation}
\CD =
\sin(\omega^{\rm (g)}/\omega + \tilde \phi^{\rm (g)})
\sin(\omega/\omega^{\rm (p)} + \tilde \phi^{\rm (p)})
+ \zeta \cos(\omega^{\rm (g)}/\omega + \tilde \phi^{\rm (g)})
\cos(\omega/\omega^{\rm (p)} + \tilde \phi^{\rm (p)}) = 0 \; ,
\label{eq:fulldisps}
\end{equation}
with%
\footnote{Since the integral in Eq.~(\ref{eq:delnu}) is dominated by
the outer layers, $\omega^{\rm (p)} \simeq 2 \Delta \nu$.}
\begin{equation}
\omega^{\rm (g)} = L \int_0^{r_1} {N \over r} \dd r \; , \quad
\omega^{\rm (p)} = \left( \int_{r_2}^R {\dd r \over c} \right)^{-1} \; .
\end{equation}
To illustrate the properties of the solution, Fig.~\ref{fig:avcross} shows
the result of increasing $\omega^{\rm (g)}$, keeping $\omega^{\rm (p)}$
fixed; for simplicity I assume that 
$\tilde \phi^{\rm (g)} = \tilde \phi^{\rm (p)} = 0$.
It is obvious that the increase in $\omega^{\rm (g)}$
leads to avoided crossings between frequencies
satisfying Eq.~(\ref{eq:pureg}), increasing with $\omega^{\rm (g)}$, and
the constant frequencies satisfying Eq.~(\ref{eq:purep}).
Details of an avoided crossing are shown in the lower panel 
of Fig.~\ref{fig:avcross}.
At the corresponding crossing of the uncoupled modes,
we have both
$\omega \equiv \omega_{\rm c} = k \pi \omega^{\rm (p)}$ and
$\omega^{\rm (g)}/ \omega_{\rm c} = n \pi$ for suitable integers
$k$ and $n$;
thus the crossing is defined by
$\omega^{\rm (g)} \equiv \omega^{\rm (g)}_{\rm c} =n k \pi^2$.
With
\begin{equation}
\omega^{\rm (g)} = \omega^{\rm (g)}_{\rm c} + \delta \omega^{\rm (g)} \; ,
\quad
\omega = \omega_{\rm c} + \delta \omega \; ,
\end{equation}
and expanding Eq.~(\ref{eq:fulldisps}) to second order in
$\delta \omega$, $\delta \omega^{\rm (g)}$ and $\zeta$ we obtain
\begin{equation}
\delta \omega^2 - 
{\omega_{\rm c} \over \omega^{\rm (g)}_{\rm c}} \delta \omega^{\rm (g)}  \,
\delta \omega
- \zeta {\omega_{\rm c}^2 \omega^{\rm (p)} \over \omega^{\rm (g)}_{\rm c}}
=0 \; ,
\end{equation}
with the solution
\begin{equation}
\delta \omega = 
{1 \over 2} {\omega_{\rm c} \over \omega^{\rm (g)}_{\rm c}} 
\delta \omega^{\rm (g)}
\pm \left[ 
{1 \over 4} \left( {\omega_{\rm c} \over \omega^{\rm (g)}_{\rm c}} 
\delta \omega^{\rm (g)}\right)^2
+ \zeta {\omega_{\rm c}^2 \omega^{\rm (p)} \over \omega^{\rm (g)}_{\rm c}}
\right]^{1/2} \;.
\label{eq:avcross}
\end{equation}
This is also shown in Fig.~\ref{fig:avcross}
for one of the avoided crossings and
clearly provides an excellent fit to the numerical solution.
The minimum separation between the two branches, for 
$\delta \omega^{\rm (g)} = 0$, is
\begin{equation}
\Delta \omega_{\rm min} = 2 \omega_{\rm c}
\left( \omega^{\rm (p)} \over \omega^{\rm (g)} \right)^{1/2} \zeta^{1/2} \; .
\end{equation}

\begin{figure}[htb]
\begin{center}
\includegraphics[width=9cm]{\fig/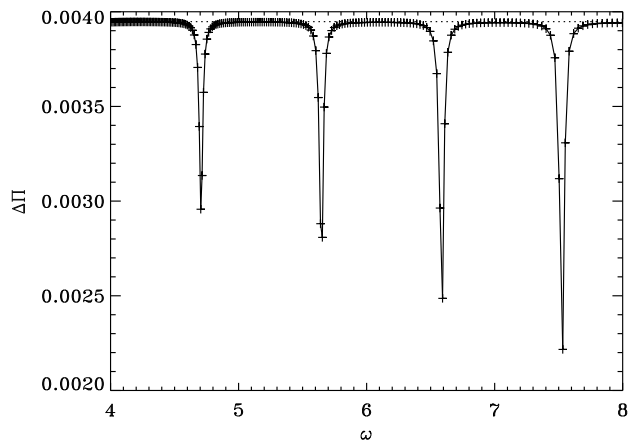}
\end{center}
\vskip -0.8cm
\caption{
Period spacings for solutions to the asymptotic dispersion relation,
Eq.~(\ref{eq:fulldisps}), with $\omega^{\rm (p)} = 0.3$,
$\omega^{\rm (g)} = 5000$ and $\zeta = 0.04$.
The period spacing $2 \pi^2 / \omega^{\rm (g)}$ for pure g modes
is shown by the horizontal dotted line.
\label{fig:asperspac}
}
\end{figure}

\begin{figure}
\begin{center}
\includegraphics[width=9cm]{\fig/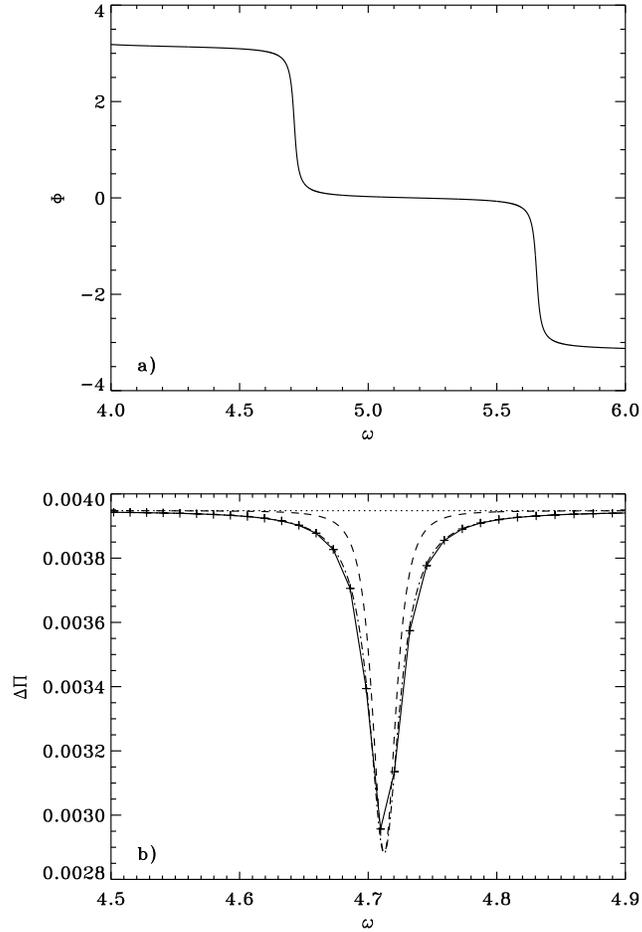}
\end{center}
\vskip -0.8cm
\caption{
Detailed properties of the case illustrated in Fig.~\ref{fig:asperspac}.
a) The frequency-dependent phase defined by Eq.~(\ref{eq:phasedef}),
to within an arbitrary multiple of $2 \pi$.
b) The pluses and solid curve show a blow-up of Fig.~\ref{fig:asperspac}.
The dot-dashed curve shows the period spacing determined from
Eq.~(\ref{eq:asperspac}), using the phase $\Phi$ shown in panel a).
The dashed curve shows the approximation in Eq.~(\ref{eq:asperspac1}).
\label{fig:asperspac1}
}
\end{figure}

As in the case of the full solution (cf.\ Fig.~\ref{fig:perspac}) 
the avoided crossings cause dips in the otherwise uniform period spacings.
This is illustrated in Fig.~\ref{fig:asperspac}.
To understand this behaviour in more detail we write Eq.~(\ref{eq:fulldisps})
as
\begin{equation}
\CD = C(\omega) \sin(\omega^{\rm (g)}/\omega + \Phi(\omega)) \; ,
\end{equation}
where (assuming again
$\tilde \phi^{\rm (g)} = \tilde \phi^{\rm (p)} = 0$)
$C(\omega) = \sqrt{\sin^2(\omega/\omega^{\rm (p)})
+ \zeta^2 \cos^2(\omega/\omega^{\rm (p)})}$ and
$\Phi(\omega)$ satisfies
\begin{eqnarray}
C(\omega) \cos(\Phi) &=& \sin(\omega / \omega^{\rm (p)}) \nonumber \\
C(\omega) \sin(\Phi) &=& \zeta \cos(\omega / \omega^{\rm (p)}) \; .
\label{eq:phasedef}
\end{eqnarray}
It is obvious that the eigenfrequencies satisfy
\begin{equation}
\CG(\omega) = \omega^{\rm (g)}/\omega + \Phi(\omega) = n \pi \; ,
\end{equation}
for integer $n$.
$\Phi$ is illustrated in Fig.~\ref{fig:asperspac1}a; 
except near acoustic resonances,
where $\omega/\omega^{\rm (p)} \simeq k \pi$ for integer $k$,
$\Phi$ is almost constant and the period spacing is determined
by the first term in $\CG$,
essentially corresponding to the pure g-mode case defined by
Eq.~(\ref{eq:pureg}).
Near the acoustic resonances $\Phi$ changes rapidly, causing a strong
variation in the period spacing.

This behaviour can be made more precise by noting that the
frequency spacing between adjacent modes approximately satisfies
$\Delta \omega \simeq \pi (\dd \CG / \dd \omega)^{-1}$ and hence the
period spacing is, approximately, given by 
\begin{equation}
\Delta \Pi \simeq - {2 \pi^2 \over \omega^2}
\left( {\dd \CG \over \dd \omega} \right)^{-1}
= {2 \pi^2 / \omega^{\rm (g)} \over
\displaystyle 1 - {\omega^2 \over \omega^{\rm (g)}}
{\dd \Phi \over \dd \omega}} \; .
\label{eq:asperspac}
\end{equation}
Here the numerator is the period spacing for pure g modes,
and the denominator causes a reduction in $\Delta \Pi$ near
an acoustic resonance.
This is illustrated in Fig.~\ref{fig:asperspac1}b where the actual
period spacings near an acoustic resonance are compared with 
Eq.~(\ref{eq:asperspac}).

At the centre of an acoustic resonance $\omega/\omega^{\rm (p)} = k \pi$
and $\Phi = \pi/2$, up to an irrelevant multiple of $\pi$.
In the vicinity of this point we can expand $\Phi$ as
$\delta \Phi = \Phi - \pi/2 $ in terms of
$\delta x = \omega/\omega^{\rm (p)} - k \pi$, to obtain
\begin{equation}
\delta \Phi \simeq - {\delta x \over \sqrt{\zeta^2 + \delta x^2}} \; ,
\end{equation}
and hence the period spacing
\begin{equation}
\Delta \Pi \simeq 
{2 \pi^2 / \omega^{\rm (g)} \over
\displaystyle 1 + {\omega^2 \over \omega^{\rm (g)} \omega^{\rm (p)}}
{\zeta^2 \over (\zeta^2 + \delta x^2)^{3/2}}} \; .
\label{eq:asperspac1}
\end{equation}
This approximation is also shown in Fig.~\ref{fig:asperspac1}b.
In particular, we obtain the minimum period spacing
\begin{equation}
\Delta \Pi_{\rm min} \simeq 
{2 \pi^2 / \omega^{\rm (g)} \over
\displaystyle 1 + {\omega^2 \over \zeta \omega^{\rm (g)} \omega^{\rm (p)}}
} \; .
\end{equation}
It is interesting that the maximum reduction in the period spacing
increases with decreasing $\zeta$ and hence weaker coupling.
On the other hand, the width of the decrease in $\Delta \Pi$ also decreases
and so therefore do the chances of finding a g mode in the vicinity
of the minimum.

Although the present analysis is highly simplified (and consequently quite
simple), it does appear to capture many of the aspects of the numerical 
solutions for stellar models. 
It would probably not be difficult to generalize it to take the behaviour 
near the turning points properly into account, 
making reasonable a more detailed comparison with the numerical results.
A comparison with the behaviour of the mode inertia (Fig.~\ref{fig:inertia})
would similarly have been interesting.
Such efforts are beyond the scope of the present paper, however.
I also note that there is clearly some symmetry between the treatment of
the g- and p-dominated modes in how the frequency-dependent phase function
$\Phi$ is introduced.
In the present case, with a dense spectrum of g modes undergoing avoided
crossings with a smaller number of p modes, it was natural to regard
the effect of the latter as a modification, in terms of $\Phi$, to the
dispersion relation for the former.
For subgiants, on the other hand, with a comparatively dense spectrum of
p modes and few g modes, a similar phase could be introduced in the 
p-mode dispersion relation. 
This might be interesting in connection with the analysis presented
by Bedding (these proceedings).

\section{Future prospects}

It is obvious that {\it Kepler} has already been an overwhelming success,
resulting in observations that will be analysed for years to come.
Even so, it is crucial to ensure that the mission continues beyond the
nominal end in late 2012.
For the exo-planet research this is required to reach the planned 
sensitivity to Earth-like planets, given a stellar background noise that
has been found to be somewhat higher than expected.
For asteroseismology the gains will be numerous.
A longer mission will allow further searches for rare types of
oscillating stars, exemplified by the recent detection in the {\it Kepler}
field of an oscillating white dwarf \citep{Ostens2011}.
The possibilities for detecting frequency variations associated with
stellar cycles in solar-like stars \citep{Karoff2009}
will be much improved.
Also, longer timeseries will allow reaching further peaks,
for modes with longer lifetimes, in the dipolar forests in
red giants and hence greatly strengthen the possibilities for investigating
the properties of their cores.

However, there is clearly a need to consider observational facilities beyond
{\it Kepler}.
The proposed ESA PLATO mission \citep{Catala2011} for transit search
for exo-planets would allow asteroseismic characterization of a large
fraction of the stars where planet systems are detected;
in addition, it would cover a much larger area on the sky than {\it Kepler},
and hence provide a greater variety of asteroseismic targets.%
\footnote{Regrettably, PLATO failed to get selected in October 2011;
it remains to be seen whether it will enter into the competition for 
a later selection.}
Also, ground-based observations of stellar oscillations in radial velocity
still present great advantages over the photometric observations 
by {\it Kepler}.
This is particularly important for solar-like stars,
where the `noise' from other processes in the stellar atmosphere provides
a background that is much higher, relative to the oscillations, in
intensity than in Doppler velocity \citep{Harvey1988, Grunda2007}.
This is the background for establishing the SONG%
\footnote{Stellar Observations Network Group}
network \citep{Christ2011b, Grunda2011} of one meter telescopes,
with a distribution to ensure almost continuous observations anywhere
in the sky.
The first node in the network, built largely with Danish funding,
will start operations in 2012, a Chinese node is under construction,
and partners are being sought to establish additional nodes.


\acknowledgements 
I wish to express my deep sympathy with the Japanese people for the losses
sustained in the disaster immediately preceding this conference.
I am very grateful to the organizers for carrying through,
in these difficult circumstances,
with the conference, with great success and in beautiful surroundings.
I thank Tim Bedding for providing Fig.~\ref{fig:obsspec},
and Travis Metcalfe and Dennis Stello for comments on earlier versions
of the manuscript.
NCAR is supported by the National Science Foundation.


\begin{thebibliography}{}

\bibitem[Aerts et al.(2010)]{Aerts2010}
Aerts, C., Christensen-Dalsgaard, J. \& Kurtz, D. W. 2010,
{\rm Asteroseismology},
(Heidelberg: Springer) 

\bibitem[Antoci et al.(2011)]{Antoci2011}
Antoci, V., Handler, G., Campante, T. L., et al. 2011,
\titnt [The excitation of solar-like oscillations in a $\delta$ Sct star
by efficient envelope convection].
{\rm Nature}, {\rm 477}, 570 

\bibitem[Baglin et al.(2009)]{Baglin2009}
Baglin, A., Auvergne, M., Barge, P., Deleuil, M., Michel, E. and
the CoRoT Exoplanet Science Team 2009,
\titnt [CoRoT: Description of the mission and early results].
in {\rm Proc. IAU Symp. 253, Transiting Planets},
edited by F. Pont, D. Sasselov and M. Holman,
(Cambridge: IAU and Cambridge University Press), 71 

\bibitem[Balmforth(1992)]{Balmfo1992}
Balmforth, N. J. 1992,
\titnt [Solar pulsational stability. I: Pulsation-mode thermodynamics].
{MNRAS}, {\rm 255}, 603 

\bibitem[Basu et al.(2011)]{Basu2011}
Basu, S., Grundahl, F., Stello, D., et al. 2011,
\titnt [Sounding open clusters: asteroseismic constraints from {\it Kepler} on
the properties of NGC 6791 and NGC 6819].
{ApJ}, {\rm 729}, L10 

\bibitem[Batalha et al.(2011)]{Batalh2011}
Batalha, N. M., Borucki, W. J., Bryson, S. T., et al. 2011,
\titnt [{\it Kepler}'s first rocky planet: Kepler-10b].
{ApJ}, {\rm 729}, 27 

\bibitem[Beck et al.(2011)]{Beck2011}
Beck, P.~G., Bedding, T.~R., Mosser, B., et al. 2011,
\titnt [Kepler detected gravity-mode period spacings in a red giant].
{\rm Science}, {\rm 332}, 205. 

\bibitem[Bedding(2011)]{Beddin2011}
Bedding, T. R. 2011,
\titnt [Solar-like oscillations: an observational perspective].
to appear in {\rm Asteroseismology},
Canary Islands Winter School of Astrophysics,
Volume XXII, edited by P. L. Pall{\'e}, (Cambridge: Cambridge University Press)
{\tt [arXiv:1107.1723]}

\bibitem[Bedding et al.(2010)]{Beddin2010}
Bedding, T.~R., Huber, D., Stello, D., et al. 2010,
\titnt [Solar-like oscillations in low-luminosity red giants: first results from
{\it Kepler}].
{ApJ}, {\rm 713}, L176 

\bibitem[Bedding \&  Kjeldsen(2003)]{Beddin2003}
Bedding, T. R. \& Kjeldsen, H. 2003,
\titnt [Solar-like oscillations].
{PASA}, {\rm 20}, 203 

\bibitem[Bedding et al.(2011)]{Beddinetal2011}
Bedding, T.~R., Mosser, B., Huber, D., et al. 2011,
\titnt [Gravity modes as a way to distinguish between hydrogen- 
and helium-burning red giant stars].
{\rm Nature}, {\rm 471}, 608 

\bibitem[Belkacem et al.(2011)]{Belkac2011}
Belkacem, K., Goupil, M. J., Dupret, M. A., Samadi, R., Baudin, F.,
Noels, A. \& Mosser, B. 2011,
\titnt [The underlying physical meaning of the $\nu_{\rm max} - \nu_{\rm c}$
relation].
{A\&A}, {\rm 530}, A142 

\bibitem[Borucki et al.(2011)]{Boruck2011}
Borucki, W. J., Koch, D. G., Basri, G., et al. 2011,
\titnt [Characteristics of planetary candidates observed by {\it Kepler}. II:
Analysis of the first four months of data].
{ApJ}, {\rm 736}, 19 

\bibitem[Borucki et al.(2009)]{Boruck2009}
Borucki, W., Koch, D., Batalha, N., Caldwell, D.,
Christensen-Dalsgaard, J., Cochran, W. D., Dunham, E., Gautier, T. N.,
Geary, J., Gilliland, R., Jenkins, J., Kjeldsen, H.,
Lissauer, J. J. \& Rowe, J. 2009,
\titnt [{\it KEPLER}: Search for Earth-size planets in the habitable zone].
in {\rm Proc. IAU Symp. 253, Transiting Planets},
edited by F. Pont, D. Sasselov and M. Holman,
(Cambridge: IAU and Cambridge University Press), 289 

\bibitem[Brown(2003)]{Brown2003}
Brown, T. M. 2003,
\titnt [Expected detection and false alarm rates for transiting Jovian planets].
{ApJ}, {\rm 593}, L125 

\bibitem[Brown et al.(1991)]{Brown1991}
Brown, T. M., Gilliland, R. L., Noyes, R. W. \& Ramsey, L. W. 1991,
\titnt [Detection of possible $p$-mode oscillations of Procyon].
{ApJ}, {\rm 368}, 599 

\bibitem[Brown et al.(2011)]{Brown2011}
Brown, T. M., Latham, D. W., Everett, M. E. \& Esquerdo, G. A. 2011,
\titnt [Kepler Input Catalog: photometric calibration and stellar classification].
{AJ}, {\rm 142}, 112 

\bibitem[Catala et al.(2011)]{Catala2011}
Catala, C., Appourchaux, T. and the PLATO Mission Consortium 2011,
\titnt [PLATO: PLAnetary Transits and Oscillations of stars].
{\rm J. Phys.: Conf. Ser.}, {\rm 271}, 012084 

\bibitem[Chaplin et al.(2005)]{Chapli2005}
Chaplin, W. J., Houdek, G., Elsworth, Y., Gough, D. O., Isaak, G. R. \&
New, R. 2005,
\titnt [On model predictions of the power spectral density of radial solar p modes].
{MNRAS}, {\rm 360}, 859 

\bibitem[Chaplin et al.(2011a)]{Chapli2011a}
Chaplin, W. J., Kjeldsen, H., Christensen-Dalsgaard, J., et al. 2011a,
\titnt [Ensemble asteroseismology of solar-type stars with the NASA Kepler mission].
{\rm Science}, {\rm 332}, 213 

\bibitem[Chaplin et al.(2011b)]{Chapli2011b}
Chaplin, W.~J., Kjeldsen, H., Bedding, T.~R., et al. 2011b,
\titnt [Predicting the detectability of oscillations in solar-type stars observed
by {\it Kepler}].
{ApJ}, {\rm 732}, 54 

\bibitem[Christensen-Dalsgaard(2004)]{Christ2004}
Christensen-Dalsgaard, J. 2004,
\titnt [Physics of solar-like oscillations].
{\rm Solar Phys.}, {\rm 220}, 137 

\bibitem[Christensen-Dalsgaard(2011a)]{Christ2011a}
Christensen-Dalsgaard, J. 2011a,
\titnt [Asteroseismology of red giants].
to appear in {\rm Asteroseismology},
Canary Islands Winter School of Astrophysics,
Volume XXII, edited by P. L. Pall{\'e}, (Cambridge: Cambridge University Press)
{\tt [arXiv:1106.5946]}

\bibitem[Christensen-Dalsgaard(2011b)]{Christ2011b}
Christensen-Dalsgaard, J. 2011b,
\titnt [Asteroseismology with Kepler and SONG].
in {\rm Proc. MEARIM II. The 2nd Middle-East and Africa Regional IAU Meeting},
edited by P. Charles,
{\rm African Skies / Cieux Africains}, in the press

\bibitem[Christensen-Dalsgaard \&  Frandsen(1983)]{Christ1983}
Christensen-Dalsgaard, J. \& Frandsen, S. 1983,
\titnt [Stellar 5 min oscillations].
{\rm Solar Phys.}, {\rm 82}, 469 

\bibitem[Christensen-Dalsgaard \&  Thompson(2011)]{ChristT2011}
Christensen-Dalsgaard, J. \& Thompson, M. J. 2011,
\titnt [Stellar hydrodynamics caught in the act: Asteroseismology with CoRoT
and Kepler].
in {\rm Proc. IAU Symposium 271: Astrophysical dynamics:
from stars to planets},
edited by N. Brummell, A. S. Brun, M. S. Miesch and Y. Ponty,
(Cambridge: IAU and Cambridge University Press), 32 

\bibitem[Christensen-Dalsgaard et al.(2008)]{Christ2008}
Christensen-Dalsgaard, J., Arentoft, T., Brown, T. M., Gilliland, R. L.,
Kjeldsen, H., Borucki, W. J. \& Koch, D. 2008,
\titnt [The {\it Kepler\/} Asteroseismic Investigation].
{\rm J. Phys.: Conf. Ser.}, {\rm 118}, 012039

\bibitem[Christensen-Dalsgaard et al.(2010)]{Christ2010}
Christensen-Dalsgaard, J., Kjeldsen, H., Brown, T.~M., Gilliland, R.~L.,
Arentoft, T., Frandsen, S., Quirion, P.-O., Borucki, W.~J., Koch, D. \&
Jenkins, J.~M. 2010,
\titnt [Asteroseismic investigation of known planet hosts in the {\it Kepler} field].
{ApJ}, {\rm 713}, L164 

\bibitem[Deheuvels \&  Michel(2011)]{Deheuv2011}
Deheuvels, S. \& Michel, E. 2011,
\titnt [Constraints on the structure of the core of subgiants via mixed modes:
the case of HD 49385].
{A\&A}, in the press
{\tt [arXiv:1109.1191v1]}

\bibitem[De Ridder et al.(2009)]{DeRidd2009}
De Ridder, J., Barban, C., Baudin, F., et al. 2009,
\titnt [Non-radial oscillation modes with long lifetimes in giant stars].
{\rm Nature}, {\rm 459}, 398 

\bibitem[Deubner \&  Gough(1984)]{Deubne1984}
Deubner, F.-L. \& Gough, D. O. 1984,
\titnt [Helioseismology: Oscillations as a diagnostic of the solar interior].
{ARA\&A}, {\rm 22}, 593 

\bibitem[Di Mauro et al.(2011)]{DiMaur2011}
Di Mauro, M.~P., Cardini, D., Catanzaro, G., et al. 2011,
\titnt [Solar-like oscillations from the depths of the red-giant star KIC 4351319
observed with {\it Kepler}].
{MNRAS}, {\rm 415}, 3783 

\bibitem[Dupret et al.(2009)]{Dupret2009}
Dupret, M.-A., Belkacem, K., Samadi, R., Montalban, J., Moreira, O.,
Miglio, A., Godart, M., Ventura, P., Ludwig, H.-G., Grigahc{\`e}ne, A.,
Goupil, M.-J., Noels, A. \& Caffau, E. 2009,
\titnt [Theoretical amplitudes and lifetimes of non-radial solar-like
oscillations in red giants].
{A\&A}, {\rm 506}, 57 

\bibitem[Frandsen et al.(2002)]{Frands2002}
Frandsen, S., Carrier, F., Aerts, C., Stello, D., Maas, T., Burnet, M.,
Bruntt, H., Teixeira, T. C., de Medeiros, J. R., Bouchy, F.,
Kjeldsen, H., Pijpers, F. \& Christensen-Dalsgaard, J. 2002,
\titnt [Detection of solar-like oscillations in the G7 giant star $\xi$ Hya].
{A\&A}, {\rm 394}, L5 

\bibitem[Gilliland et al.(2010)]{Gillil2010}
Gilliland, R. L., Brown, T. M., Christensen-Dalsgaard, J., et al. 2010,
\titnt [{\it Kepler} asteroseismology program: Introduction and first results].
{PASP}, {\rm 122}, 131 

\bibitem[Goldreich \&  Keeley(1977)]{Goldre1977}
Goldreich, P. \& Keeley, D. A. 1977,
\titnt [Solar seismology. II. The stochastic excitation of the solar
$p$-modes by turbulent convection].
{ApJ}, {\rm 212}, 243 

\bibitem[Gough(1993)]{Gough1993}
Gough, D. O. 1993,
\titnt [Course 7. Linear adiabatic stellar pulsation].
in {\rm Astrophysical fluid dynamics, Les Houches Session XLVII},
edited by J.-P. Zahn and J. Zinn-Justin, (Amsterdam: Elsevier),  399 

\bibitem[Gough(2007)]{Gough2007}
Gough, D. O. 2007,
\titnt [An elementary introduction to the JWKB theory].
{AN}, {\rm 328}, 273 

\bibitem[Grundahl et al.(2007)]{Grunda2007}
Grundahl, F., Kjeldsen, H., Christensen-Dalsgaard, J., Arentoft, T. \&
Frandsen, S. 2007,
\titnt [Stellar Oscillations Network Group].
{\rm Comm. in Asteroseismology}, {\rm 150}, 300 

\bibitem[Grundahl et al.(2011)]{Grunda2011}
Grundahl, F., Christensen-Dalsgaard, J., J{\o}rgensen, U. G., Kjeldsen, H.,
Frandsen, S. \& Kj{\ae}rgaard Rasmussen, P. 2011,
\titnt [SONG -- getting ready for the prototype].
{\rm J. Phys.: Conf. Ser.}, {\rm 271}, 012083 

\bibitem[Harvey(1988)]{Harvey1988}
Harvey, J. W. 1988,
\titnt [Techniques for observing stellar oscillations].
in {\rm Proc. IAU Symposium No 123, Advances in helio- and asteroseismology},
edited by J. Christensen-Dalsgaard and S. Frandsen, 
(Dordrecht: Reidel), 497 

\bibitem[Hekker et al.(2009)]{Hekker2009}
Hekker, S., Kallinger, T., Baudin, F., De Ridder, J., Barban, C.,
Carrier, F., Hatzes, A. P., Weiss, W. W. \& Baglin, A. 2009,
\titnt [Characteristics of solar-like oscillations in red giants observed in the
CoRoT exoplanet field].
{A\&A}, {\rm 506}, 465 

\bibitem[Hekker et al.(2011)]{Hekker2011}
Hekker, S., Gilliland, R. L., Elsworth, Y., Chaplin, W. J., De Ridder, J.,
Stello, D., Kallinger, T., Ibrahim, K. A., Klaus, T. C. \& Li, J. 2011,
\titnt [Characterisation of red giant stars in the public {\it Kepler} data].
{MNRAS},  {\rm 414}, 2594 

\bibitem[Houdek et al.(1999)]{Houdek1999}
Houdek, G., Balmforth, N. J., Christensen-Dalsgaard, J. \& Gough, D. O. 1999,
\titnt [Amplitudes of stochastically excited oscillations in main-sequence stars].
{A\&A}, {\rm 351}, 582 

\bibitem[Huber et al.(2010)]{Huber2010}
Huber, D., Bedding, T. R., Stello, D., et al. 2010,
\titnt [Asteroseismology of red giants from the first four months of 
{\it Kepler} data: global oscillation parameters for 800 stars].
{ApJ}, {\rm 723}, 1607 

\bibitem[Huber et al.(2011)]{Huber2011}
Huber, D., Bedding, T.~R., Stello, D., et al. 2011,
\titnt [Testing scaling relations for solar-like oscillations from 
the main-sequence to red giants using {\it Kepler} data].
{ApJ}, in the press 
{\tt [arXiv:1109.3460]}

\bibitem[Jiang et al.(2011)]{Jiang2011}
Jiang, C., Jiang, B. W., Christensen-Dalsgaard, J., et al. 2011,
{ApJ}, in the press 
{\tt [arXiv:1109.0962]}

\bibitem[Kallinger et al.(2010)]{Kallin2010}
Kallinger, T., Mosser, B., Hekker, S., et al. 2010,
\titnt [Asteroseismology of red giants from the first four months of {\it Kepler}
data: Fundamental stellar parameters].
{A\&A}, {\rm 522}, A1 

\bibitem[Karoff et al.(2009)]{Karoff2009}
Karoff, C., Metcalfe, T. S., Chaplin, W. J., Elsworth, Y., Kjeldsen, H.,
Arentoft, T. \& Buzasi, D. 2009,
\titnt [Sounding stellar cycles with Kepler -- I. Strategy for selecting targets].
{MNRAS}, {\rm 399}, 914 

\bibitem[Kjeldsen et al.(2010)]{Kjelds2010}
Kjeldsen, H., Christensen-Dalsgaard, J., Handberg, R., Brown, T. M., 
Gilliland, R. L., Borucki, W. J. \& Koch, D. 2010,
\titnt [The {\it Kepler} Asteroseismic Investigation: Scientific goals and the
first results].
{AN}, {\rm 331}, 966 

\bibitem[Koch et al.(2010)]{Koch2010}
Koch, D. G., Borucki, W. J., Basri, G., et al. 2010,
\titnt [{\it Kepler Mission} design, realized photometric performance, and
early science].
{ApJ}, {\rm 713}, L79 

\bibitem[Lamb(1909)]{Lamb1909}
Lamb, H. 1909,
\titnt [On the theory of waves propagated vertically in the atmosphere].
{\rm Proc. London Math. Soc.}, {\rm 7}, 122 

\bibitem[L{\'e}ger et al.(2009)]{Leger2009}
L{\'e}ger, A., Rouan, D., Schneider, J., et al. 2009,
\titnt [Transiting exoplanets from the CoRoT space mission. VIII. CoRoT-7b: 
the first super-Earth with measured radius].
{A\&A}, {\rm 506}, 287 

\bibitem[Lissauer et al.(2011)]{Lissau2011}
Lissauer, J. J., Fabrycky, D. C., Ford, E. B., et al. 2011,
\titnt [A closely packed system of low-mass, low-density planets transiting
Kepler-11].
{\rm Nature}, {\rm 470}, 53 

\bibitem[Metcalfe et al.(2010)]{Metcal2010}
Metcalfe, T. S., Monteiro, M. J. P. F. G., Thompson, M. J., et al. 2010,
\titnt [A precise asteroseismic age and radius for the evolved Sun-like star
KIC 11026764].
{ApJ}, {\rm 723}, 1583 

\bibitem[Michel et al.(2008)]{Michel2008}
Michel, E., Baglin, A., Auvergne, M., et al. 2008,
\titnt [CoRoT measures solar-like oscillations and granulation in stars hotter
than the Sun].
{\rm Science}, {\rm 322}, 558 

\bibitem[Miglio et al.(2009)]{Miglio2009}
Miglio, A., Montalb{\'a}n, J., Baudin, F., Eggenberger, P., Noels, A., 
Hekker, S., De Ridder, J., Weiss, W. \& Baglin, A. 2009,
\titnt [Probing populations of red giants in the galactic disk with CoRoT].
{A\&A}, {\rm 503}, L21 

\bibitem[Miglio et al.(2011)]{Miglio2011}
Miglio, A., Brogaard, K., Stello, D., et al. 2011,
\titnt [Asteroseismology of old open clusters with Kepler: direct estimate of the integrated RGB mass loss in NGC6791 and NGC6819].
{MNRAS}, in the press
{\tt [arXiv:1109.4376]}.

\bibitem[Molenda-{\.Z}akowicz et al.(2010)]{Molend2010}
Molenda-{\.Z}akowicz, J., Bruntt, H., Sousa, S., et al. 2010,
\titnt [Asteroseismology of solar-type stars with Kepler: III. Ground-based data].
{AN}, {\rm 331}, 981 

\bibitem[Molenda-\.Zakowicz et al.(2011)]{Molend2011}
Molenda-\.Zakowicz, J., Latham, D. W., Catanzaro, G., Frasca, A. \&
Quinn, S. N. 2011,
\titnt [Characterizing {\it Kepler} asteroseismic targets].
{MNRAS}, {\rm 412}, 1210 

\bibitem[Montalb{\'a}n et al.(2010)]{Montal2010}
Montalb{\'a}n, J., Miglio, A., Noels, A., Scuflaire, R. \& Ventura, P. 2010,
\titnt [Seismic diagnostics of red giants: first comparison with stellar models].
{ApJ}, {\rm 721}, L182 

\bibitem[Mosser et al.(2010)]{Mosser2010}
Mosser, B., Belkacem, K., Goupil, M.-J., et al. 2010,
\titnt [Red-giant seismic properties analyzed with CoRoT].
{A\&A}, {\rm 517}, A22 

\bibitem[Mosser et al.(2011a)]{Mosser2011a}
Mosser, B., Belkacem, K., Goupil, M. J., et al. 2011a,
\titnt [The universal red-giant oscillation pattern. An automated determination 
with CoRoT data].
{A\&A}, {\rm 525}, L9 

\bibitem[Mosser et al.(2011b)]{Mosser2011b}
Mosser, B., Elsworth, Y., Hekker, S., et al. 2011b,
\titnt [Characterization of the power excess of solar-like oscillations in
red giants with Kepler].
{A\&A}, in the press
{\tt [arXiv:1110.0980v1]}

\bibitem[Mosser et al.(2011c)]{Mosser2011c}
Mosser, B., Barban, C., Montalb{\'a}n, J., et al. 2011c,
\titnt [Mixed modes in red-giant stars observed with CoRoT].
{A\&A}, {\rm 532}, A86 

\bibitem[{\O}stensen et al.(2011)]{Ostens2011}
{\O}stensen, R. H., Bloemen, S., Vu{\v c}kovi{\'c}, M., Aerts, C.,
Oreiro, R., Kinemuchi, K., Still, M. \& Koester, D. 2011,
\titnt [At last---a V777\,Her pulsator in the {\it Kepler} field].
{ApJ}, {\rm 736}, L39 

\bibitem[Queloz et al.(2009)]{Queloz2009}
Queloz, D., Bouchy, F., Moutou, C., et al. 2009,
\titnt [The CoRoT-7 planetary system: two orbiting super-Earths].
{A\&A}, {\rm 506}, 303 

\bibitem[Stello et al.(2008)]{Stello2008}
Stello, D., Bruntt, H., Preston, H. \& Buzasi, D. 2008,
\titnt [Oscillating K giants with the {\it WIRE\/} satellite: 
determination of their asteroseismic masses].
{ApJ}, {\rm 674}, L53 

\bibitem[Stello et al.(2010)]{Stello2010}
Stello, D., Basu, S., Bruntt, H., et al. 2010,
\titnt [Detection of solar-like oscillations from {\it Kepler} photometry 
of the open cluster NGC 6819].
{ApJ}, {\rm 713}, L182 

\bibitem[Stello et al.(2011a)]{Stello2011a}
Stello, D., Meibom, S., Gilliland, R. L., et al. 2011a,
\titnt [An asteroseismic membership study of the red giants in three open clusters
observed by {\it Kepler}: NGC 6791, NGC 6819, and NGC 6811].
{ApJ}, {\rm 739}, 13 

\bibitem[Stello et al.(2011b)]{Stello2011b}
Stello, D., Huber, D., Kallinger, T., et al. 2011b,
\titnt [Amplitudes of solar-like oscillations: constraints from red giants in
open clusters observed by {\it Kepler}].
{ApJ}, {\rm 737}, L10 

\bibitem[Tassoul(1980)]{Tassou1980}
Tassoul, M. 1980,
\titnt [Asymptotic approximations for stellar nonradial pulsations].
{ApJS}, {\rm 43}, 469 

\bibitem[Unno et al.(1989)]{Unno1989}
Unno, W., Osaki, Y., Ando, H., Saio, H. \& Shibahashi, H. 1989,
{\rm Nonradial Oscillations of Stars, 2nd Edition}
(Tokyo: University of Tokyo Press)

\bibitem[Uytterhoeven et al.(2010)]{Uytter2010}
Uytterhoeven, K., Briquet, M., Bruntt, H., et al. 2010,
\titnt [Ground-based follow-up in relation to Kepler Asteroseismic Investigation].
{AN}, {\rm 331}, 993 

\bibitem[Vandakurov et al.(1967)]{Vandak1967}
Vandakurov, Yu. V. 1967,
\titnt [The frequency distribution of stellar oscillations].
{AZh}, {\rm 44}, 786 
(English translation: {\rm Soviet Ast.}, {\rm 11}, 630)

\bibitem[Walker et al.(2003)]{Walker2003}
Walker, G., Matthews, J., Kuschnig, R., et al. 2003,
\titnt [The {\it MOST} asteroseismology mission: Ultraprecise photometry from Space].
{PASP}, {\rm 115}, 1023 

\bibitem[White et al.(2011)]{White2011}
White, T. R., Bedding, T. R., Stello, D., Christensen-Dalsgaard, J.,
Huber, D. \& Kjeldsen, H. 2011,
\titnt [Calculating asteroseismic diagrams for solar-like oscillations].
{ApJ}, in the press 
{\tt [arXiv:1109.3455]}

\bibitem[Winn et al.(2011)]{Winn2011}
Winn, J. N., Matthews, J. M., Dawson, R. I., et al. 2011,
\titnt [A super-Earth transiting a naked-eye star].
{ApJ}, {\rm 737}, L18 

\end{thebibliography}

\end{document}